\documentclass[aps,12pt]{revtex4}
\usepackage[latin1]{inputenc}
\usepackage[dvips]{color}
\usepackage{graphicx} 
\usepackage{epstopdf}
\usepackage{amsfonts,amstext,amsbsy,amssymb,amsmath}

\usepackage[colorlinks = true,allcolors = {blue}]{hyperref}

\renewcommand{\Re}{Re}
\newcommand{\Ra}{Ra}
\newcommand{\Ta}{{Ta}}

\newcommand{\Nu}{Nu}
\newcommand{\Nuw}{Nu_{\omega}}

\newcommand{\lam}{\operatorname{lam}}
\newcommand{\Ro}{Ro}

\begin{document}

\title{Rough wall turbulent Taylor-Couette flow: \\ the effect of the rib height}
\author{Ruben A. Verschoof$^1$,
  Xiaojue Zhu$^1$,
  Dennis Bakhuis$^1$,
  Sander G. Huisman$^1$,
  Roberto Verzicco$^{2,1}$
 Chao Sun$^{3,1}$\footnote{chaosun@tsinghua.edu.cn},
and Detlef Lohse$^{1,3,4}$\footnote{d.lohse@utwente.nl} 
 }
\affiliation{
$^1$Physics of Fluids, Max Planck Institute for Complex Fluid Dynamics, MESA+ institute and J. M. Burgers Center for Fluid Dynamics, University of Twente, P.O. Box 217, 7500 AE Enschede, The Netherlands\\
$^2$ Dipartimento di Ingegneria Industriale, University of Rome ``Tor Vergata'', Via del Politecnico 1, Roma 00133, Italy\\
$^3$Center for Combustion Energy and Department of Energy and Power Engineering, Tsinghua University, 100084 Beijing, China\\
$^4$Max Planck Institute for Dynamics and Self-Organization, 37077 G\"{o}ttingen, Germany}
\date{\today}

\begin{abstract}
In this study, we combine experiments and direct numerical simulations to investigate the effects of the height of transverse ribs at the walls on both global and local flow properties in turbulent Taylor-Couette flow. We create rib roughness by attaching up to 6 axial obstacles to the surfaces of the cylinders over an extensive range of rib heights, up to blockages of 25\% of the gap width. In the asymptotic ultimate regime, where the transport is independent of viscosity, we emperically find that the prefactor of the $\Nuw \propto \Ta^{1/2}$ scaling (corresponding to the drag coefficient  $C_f(\Re)$ being constant) scales with the number of ribs $N_r$ and by the rib height $h^{1.71}$. The physical mechanism behind this is that the dominant contribution to the torque originates from the pressure forces acting on the rib which scale with rib height. The measured scaling relation of $N_r h^{1.71}$ is slightly smaller than the expected $N_r h^2$ scaling, presumably because the ribs cannot be regarded as completely isolated but interact.
In the counter-rotating regime with smooth walls, the momentum transport is increased by turbulent Taylor vortices. We find that also in the presence of transverse ribs these vortices persist. In the counter-rotating regime, even for large roughness heights, the momentum transport is enhanced by these vortices.

\end{abstract}

\maketitle

\vspace{8mm}

\section{Introduction}
Turbulent flows with rough walls are omnipresent in nature and industry. In fact, for increasing Reynolds numbers, the viscous length-scales in the flow decrease, and eventually every surface appears to be rough even when the roughness is small in absolute scale. Roughness in turbulent flows is relevant in many fields, one can think of e.g.\ biofouling in marine vessels \cite{sch07}, atmospheric boundary layers \cite{ren11}, and the accelerated transition to turbulence, see e.g.\ refs.\ \cite{mar10b,fla14} for recent reviews.
 The study of roughness in turbulent flows has received tremendous attention, especially the field of rough pipe flow studies has a long history. The most seminal work to date remains the well-known pipe flow experiments by Nikuradse \cite{nik33}. He expressed the friction as a dimensionless friction factor $C_f$, and found that $C_f$ decreases for increasing Reynolds number $\Re$, and eventually becomes constant in the presence of roughness. The absolute value of $C_f$ then depends on the characteristic height of the roughness. Using that work and successive work of Colebrook \cite{Colebrook1939} and Moody \cite{moo44}, engineers were enabled to estimate the pressure drop in pipes. 
However, the influence of roughness on turbulent flows remains far from being understood. Many experimental studies focussed on industrial applicability, and have not emphasized the physical understanding of the flow dynamics, as was pointed out by ref.\ \cite{jim04}. Furthermore, roughness remains hard to quantify given the huge variety of roughness types. Although significant progress has been made in recent years, the study of roughness in highly turbulent flows remains a topic of great interest to both physicists and engineers \cite{Shockling2006,Schultz2007,busse2015,cha15,MacDonald2016,Thakkar2018, Flack2018}.

In this study, building further upon our recent work \cite{zhu18}, we use a Taylor-Couette system, i.e.\ the fluid flow between two concentric, independently rotating cylinders, to study the effects of transverse ribs in highly turbulent flow. Taylor-Couette (TC) flow has the advantages of (i) being a closed flow with an exact balance between energy input and dissipation, (ii) being accesible to study both numerically and experimentally due to its simple geometry and high symmetries, (iii) having no streamwise spatial transients and (iv) being mathematically well-defined based on the Navier-Stokes equations, the continuity equation and the known boundary conditions.

Indeed, Taylor-Couette flow, together with pipe flow and Rayleigh-B\'enard (RB) convection, is one of the canonical systems in which the physics of fluids is studied \cite{bus12,far14,gro16}.  When the correct dimensionless parameters are used, the scaling relations between driving and response are the same for RB convection and TC flow, namely \cite{gro11}
\begin{equation}
\Nuw \propto \Ta^{1/2}\mathcal{L}(\Re),~~~~ \text{and}~~~~ \Nu \propto \Ra^{1/2}\mathcal{L}(\Re), 
\end{equation}
for fixed (geometric) Prandtl number. The terms $\mathcal{L}(Re)$ are logarithmic corrections, which are related to a viscosity-dependence in the turbulent boundary layers \cite{kra62,gro11}. For smooth walls, effective power laws of $\Nuw\propto Ta^{0.38}$ and $\Nu \propto \Ra^{0.38}$ are found both numerically and experimentally in TC flow and RB convection in hitherto studied parameter regions ($10^8 \leq \Ta  \leq 10^{13}$ and $5 \times 10^{14} \leq \Ra \leq 10^{15}$)  \cite{gil11,he12a,he12,ost14pd}. 

Recently, building on prior work \cite{cad97,ber03} we showed that attaching ribs on both cylinders in a TC setup is an effective way to attain a $\Nuw \propto \Ta^{1/2}$ scaling without log-corrections \cite{zhu18}. Thanks to the ribs, the viscosity-dependence in the boundary layers is eliminated as the dissipative process becomes fully pressure-dominated. This scaling, which we called ``asymptotic ultimate turbulence scaling'' has the same scaling as  the mathematical upper bound of momentum transport \cite{doe96,nic97}, namely $\Nuw \propto \Ta^{1/2}$, where the prefactor of the $\Nuw \propto \Ta^{\gamma}$ scaling still depends on the roughness characteristics, i.e.\ the height and number of ribs. Similar observations were made in pipe flow \cite{Webb1971}, as $\Nuw \propto \Ta^{1/2}$ is mathematically equivalent to having a Reynolds number independent drag coefficient  $C_f$. In fact, what we here refer to as ``asymptotic ultimate turbulence''  is identical to the so-called ``fully rough'' regime in pipe flow.

The geometry of a Taylor-Couette setup is characterized by two geometric parameters, which are the radius ratio $\eta = r_i/r_o$ and the aspect ratio $\Gamma = L / (r_o-r_i)$, in which $r_i$ and $r_o $ are the radii of the inner and outer cylinder, respectively, and $L$ is the height of the setup. Taylor-Couette flow is driven by the rotation of one or both cylinders. Their driving can be expressed as two different Reynolds numbers, i.e. 
\begin{equation}
\Re_i = \frac{\omega_i r_i d}{\nu},~~~~\text{and}~~~~ \Re_o = \frac{\omega_o r_o d}{\nu},
\end{equation}
in which $\nu$ is the kinematic viscosity, $d=r_o-r_i$ is the gap width and $\omega_{i}$ and $\omega_o$ are the angular velocities of the inner and outer cylinders, respectively. Alternatively, we can express the driving using the Taylor number $\Ta$ and a rotation ratio. The Taylor number,  being equivalent to the Rayleigh number in RB convection, is given as
\begin{equation}
\Ta = \frac{\sigma}{4}\frac{ d^2(r_i^2+r_o^2)(\omega_i - \omega_o)^2}{\nu^2} \propto (\Re_i - \eta \Re_o)^2.
\end{equation}
Here $\sigma=\left(\frac{1+\eta}{2\sqrt{\eta}}\right)^4$, which is a fixed geometric parameter, is referred to as a `geometric Prandtl number' \cite*{eck07b}. The rotation ratio between both cylinders is given as $a=-\omega_o / \omega_i$, and can also be expressed in terms of an inverse Rossby number, which directly enters the equations of motion as the Coriolis force, as will be discussed in section \ref{subsec:numerics}. The rotation ratio $a$ and the inverse Rossby number are related by
\begin{equation}
\Ro^{-1} =\frac{2 \omega_o d}{|\omega_i - \omega_o| r_i} =- 2 \frac{a}{|1+a|}\frac{1-\eta}{\eta}. 
\end{equation}

The primary response parameter is the torque $\tau$ necessary to rotate the cylinders at a given driving  $Re_i$ and $Re_o$, or, equivalently, $Ta$ and $a$. To underline the analogy with RB convection, the torque is expressed as the Nusselt number, which is the ratio between the angular velocity flux $J^{\omega}$ and its laminar value $J^{\omega}_{lam} = 2 \nu r_i^2r_o^2(\omega_i - \omega_o) / (r_o^2-r_i^2)$. The Nusselt number is directly related to the torque by 
\begin{equation}
\Nuw = J^{\omega} / J^{\omega}_{lam} = \frac{\tau}{2 \pi L \rho J_{\lam}^{\omega}},
\end{equation}
in which $\rho$ is the density of the fluid. These equations hold for all cases, including co- and counter-rotation, and are also valid when ribs are added to the cylinders. A different nondimensional representation of the torque is  as friction coefficient $C_f$, which is traditionally used in the  wall-bounded turbulence community
\begin{equation}
C_f = \frac{\tau}{L \rho \nu^2  (\Re_i - \eta \Re_o)^2}.
\end{equation}
Lastly, the torque can be related to the friction velocity at either the inner or the outer cylinder $u_{\tau} = \sqrt{\tau/2 \pi \rho r_{i,o}^2 L}$, which can be used, along with the viscous lengthscale $\delta_{\nu} = \nu/u_{\tau}$, to express the velocity and wall-normal distance in wall units.

In this work, we build further on our prior work \cite{zhu18}, by combining detailed experiments and direct numerical simulations, to quantify the influence of the rib height on the prefactors of the scaling relations. Furthermore, we study the role of the pressure acting on the ribs, as well as the local flow response. 

The outline of this article is as follows; We first discuss the experimental and numerical methods, as well as the explored parameter space in section \ref{sec:methods}. We continue with presenting global flow results in section \ref{sec:global}, which is followed by local flow results in section \ref{sec:local}. In section \ref{sec:CR}, we explore the regime of counter-rotating cylinders. We conclude this manuscript in section \ref{sec:conc}.

%% Omrekenen CF -> Nu/Ta^0.5
\iffalse
\begin{align}
c_f &= \frac{\tau}{L \rho \nu^2 (Re_i - \eta Re_o)^2} = \frac{\tau}{L \rho d^2 r_i^2(\omega_i -  \omega_o)^2}\\
Ta^{1/2} &= \frac{\sqrt{\sigma}}{2} d \sqrt{r_i^2+r_o^2}(\omega_i - \omega_o) /\nu\\
\Nuw &= \frac{\tau(r_o^2 - r_i^2)}{4 \pi \rho L \nu r_i^2 r_o^2 (\omega_i - \omega_o)}\\
\Nuw/\Ta^{1/2} &= \frac{\tau(1-\eta^2)}{2 \pi \rho L r_i^2(\omega_i-\omega_o)^2\sqrt{\sigma}d \sqrt{r_i^2+ r_o^2}}\\
& = \frac{\tau(1-\eta^2)}{\rho L d r_i^2  (\omega_i-\omega_o)^2 2 \pi \sqrt{\sigma} \sqrt{r_i^2+ r_o^2}} \\
& = c_f \frac{d(1-\eta^2)}{2 \pi \sqrt{\sigma} \sqrt{r_i^2+ r_o^2}}\\
& = c_f \frac{(1-\eta)(1-\eta^2)}{2 \pi \sqrt{\sigma} \sqrt{\eta^2+1}}\\
& = c_f 0.0173
\end{align}
\fi

\section{Methods}
\label{sec:methods}
\subsection{Experimental methods}

The experiments were performed in the Twente Turbulent Taylor-Couette (T$^3$C) facility \cite{gil11a}. The setup has an inner cylinder with a radius of $r_i=$ 200.0 mm and an outer cylinder with a radius of $r_o=$ 279.4 mm, resulting in a radius ratio of $\eta = r_i/r_o = 0.716$ and a gap width of $d=r_o-r_i=$ 79.4 mm. The gap is filled with water at a temperature of T $ = 20 \pm 0.5 ^{\circ}$C, which is kept constant by active cooling through the top and bottom plates. Nonetheless, the temperature is monitored continuously, such that the viscosity is calculated using the instantaneous temperature. 
In this work, the inner and outer cylinder rotate up to $\omega_i/(2\pi)=$ 10 Hz and up to $\omega_o/(2\pi)= \pm$ 5 Hz, respectively, resulting in Reynolds numbers up to $\Re_i = \omega_i r_i d/\nu = 1 \times 10^6$ and $\Re_o= \omega_o r_o d/\nu = 7 \times 10^5$. The cylinders have a height of $L =$ 927 mm, resulting in an aspect ratio of $\Gamma = L/(r_o-r_i) = 11.7$. The end plates rotate with the outer cylinder.

\begin{figure}
\centering
\includegraphics[scale=1.2]{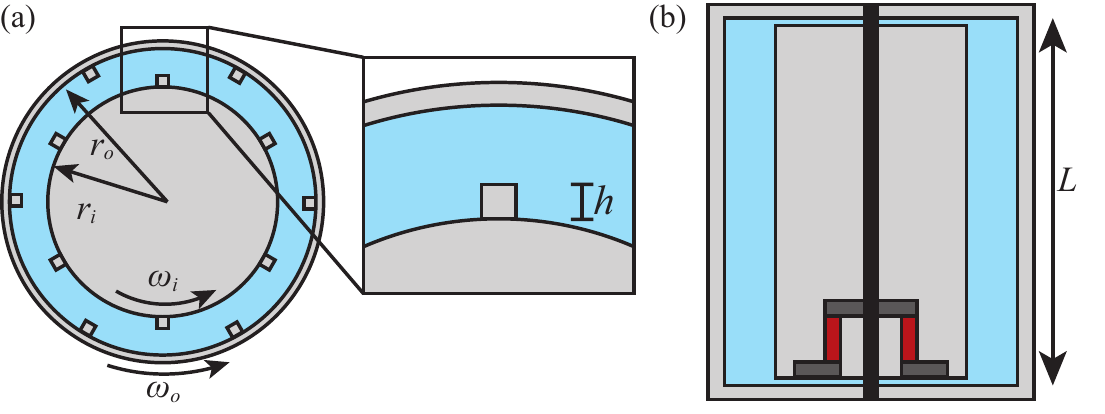}
\caption{{\bf Schematic of the experimental setup}  {\bf (a)} Top view of the experimental setup, in which ribs (not to scale) are placed on both the inner and outer cylinder. The ribs extend over the entire height of the cylinders. The zoom shows how the rib height $h$ is defined. {\bf (b)} Cross-section of the TC setup. The torque sensor is located in the inner cylinder. }
\label{Chap_Four_fig:fig1}
\end{figure}

The torque is measured with a co-axial torque transducer (Honeywell 2404-1K, maximum capacity of 115 Nm), located inside the inner cylinder to avoid measurement errors due to friction of seals and bearings, as shown in figure\ \ref{Chap_Four_fig:fig1}. In previous studies using this setup, the inner cylinder consisted of 3 different sections, and the torque was measured only in the middle section to reduce end plate effects \cite{gil11,gil12}. Here, we measure over the entire height of the cylinder, which accounts for the slightly different results for the smooth-wall case as compared to those studies.

\subsection{Numerical methods}
\label{subsec:numerics}
We numerically solve the incompressible Navier-Stokes equations in the  frame of reference which co-rotates with the outer cylinder
\begin{eqnarray}
\label{equ1}
\frac{\partial \bf{u}}{\partial t}+{\bf{u}} \cdot \nabla {\bf u} &=&- \nabla p + \frac{f(\eta)}{\Ta^{1/2}}{\nabla}^2 {\bf u}-\Ro^{-1}{\bf e }_z\times {\bf u }, \\
\nabla  \cdot {\bf u} &=& 0, 
\end{eqnarray}
where $\bf u$ is the fluid velocity,  $p$ the pressure, and ${\bf e }_z$ the unit vector in the axial direction. $f(\eta)$ is a geometrical factor which has the form 
\begin{eqnarray}
f(\eta)= \frac{(1+\eta)^3}{8\eta^2}.
\end{eqnarray}
The direct numerical simulations were carried out by solving the above governing equations, using a second order finite difference code AFiD \cite{ver96,poe15cf}, in combination with an immersed-boundary method \cite{fad00,yan06} for the rotating roughness elements. A two-dimensional MPI decomposition technique (MPI-pencil) \cite{poe15cf} was implemented to achieve highly parallelized computation. In recent years, we have tested the code extensively for TC flow with smooth \cite{ost14pof, ost14pd, ost15pof} and rough \cite{zhu17,zhu18} walls. The boundary condition in the axial direction is periodic and thus we do not have end plate effects, which, as ref.\ \cite{avi12} showed, are small in the turbulent regime. The radius ratio is chosen as $\eta=0.716$, the same as in experiments. The aspect ratio of the computational domain $\Gamma=L/d$, where $L$ is the axial periodicity length, is taken as $\Gamma=2.09$. The azimuthal extent of the domain is $\pi/3$, to reduce computation costs without affecting the results \cite{bra13}. The computation box is tested to be large enough to capture the sign changes of the azimuthal velocity autocorrelation at the mid-gap, which was suggested as a criterion for the box size \cite{ost15pof}. For more information on the numerical details, we refer to ref.\ \cite{zhu18}. 

\subsection{Explored parameter space}
Experimentally, we explored Taylor numbers of $O(10^{10})$ to $O(10^{13})$. Numerically, all Taylor numbers below $O(10^{10})$ are accessible. The exact values depend on the roughness size. In that sense, the simulations  and experiments are completely complementary, and we explore a parameter space in which $Ta$ extends over 5 orders of magnitude. 
We restrict ourselves to, (i) equidistant ribs, and (ii) the same number of ribs on both the inner and outer cylinder.
Numerically, we attach 6 ribs to both cylinders. The used rib heights are 1.5\%, 2.5\%, 5\%, 7.5\% and 10\% of the gap width. The maximum blockage obviously is twice as large: e.g.\ with when 2 ribs with a 10\% rib height pass, the local blockage is $20\%$ of the gap width.

Experimentally, we attach 2, 3, or 6 vertical ribs to both cylinders, as shown in figure\ \ref{Chap_Four_fig:fig1}. The used roughness heights are 2 mm, 4 mm, 6 mm, 8 mm, and 10 mm, corresponding to $2.5\%$, $5\%$, $7.5\%$, $10\%$, and $12.5\%$ of the gap width. For both the simulations and experiments, we also measure without ribs as reference smooth-wall case. Numerically, we restricted ourselves to inner cylinder rotation only, whereas experimentally we also explore the counter-rotating regime.

\section{ Global response: torque and its scaling}
\label{sec:global}

\begin{figure}
\centering
\includegraphics[scale=1]{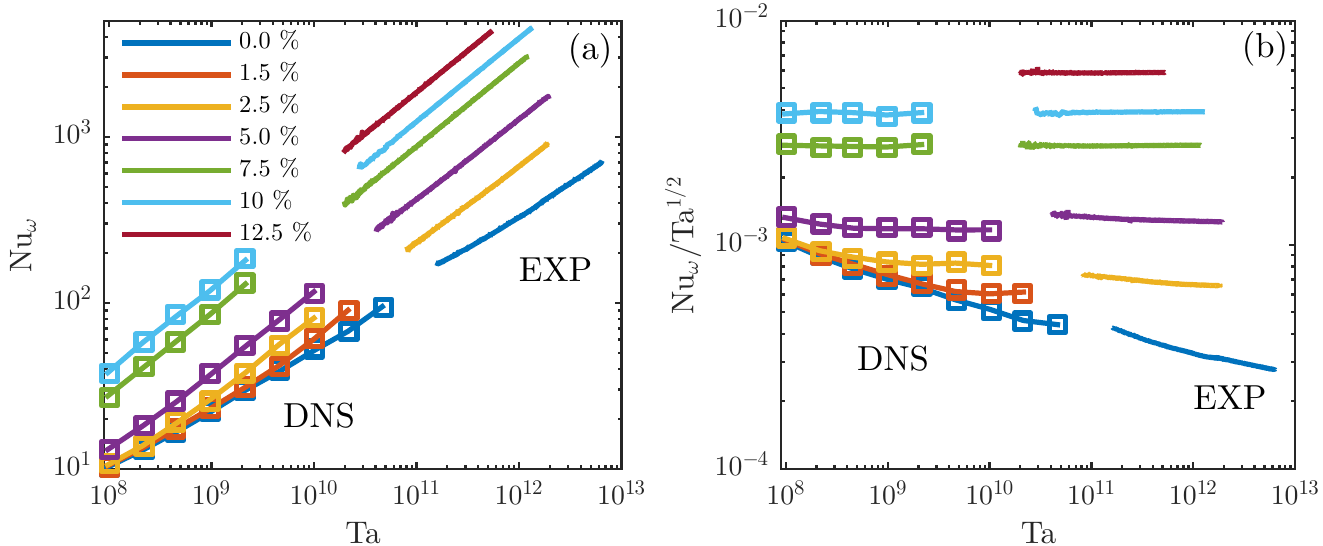}
\caption{Torque scaling as a function of driving for  $a=0$, i.e.\ pure inner cylinder rotation. 6 ribs are attached to each cylinders. As indicated in the figure, the lower Taylor number data points are DNSs, the higher Taylor number data are experimental results. {\bf (a)}   The Nusselt number $\Nuw$ as a function of Taylor number $\Ta$.   {\bf (b)} Here we compensate $\Nuw$ with $\Ta^{1/2}$ to reveal whether the asymptotic ultimate regime is reached. The local scaling depends on both the Taylor number and the roughness height.}
\label{Chap_Four_fig:fig2}
\end{figure}

The global response of momentum transport in TC flow can be expressed as the torque which is necessary to keep the cylinders rotating at fixed angular velocities. Here we show the dimensionless torque $\Nuw$ as a function of driving, expressed here as the Taylor number $Ta$. In figure \ref{Chap_Four_fig:fig2} we show results for 6 ribs on both cylinders for the case with a stationary outer cylinder. This figure clearly shows that $\Nuw$ is increased {\it tremendously} as the roughness height increases. To highlight the local scaling, we compensate $\Nuw$ with $\Ta^{1/2}$ in figure \ref{Chap_Four_fig:fig2}b. This figure is very similar to the well-known Moody diagram for pipe flow, in which the friction coefficient $C_f$ is expressed as a function of the Reynolds number \cite{moo44,zhu18}. In fact, using the definitions given above, and apart from a prefactor, $\Nuw/ \Ta^{1/2}$ and $C_f$ are identical, as they are related as 
\begin{equation}
\frac{\Nuw}{ \Ta^{1/2}} = \frac{(1-\eta)(1-\eta^2)}{2 \pi \sqrt{\sigma}\sqrt{\eta^2+1}} C_f = \frac{2(\eta-1)^2\eta}{\pi (1+\eta)\sqrt{1+\eta^2}} C_f.
\end{equation}
 For the currently used radius ratio of $\eta=0.716$, this results in  $\Nuw/ \Ta^{1/2} =0.0174 C_f$.
Here, we see that for the currently studied Taylor number regime, all cases with ribs larger than $7.5\%$ of the gap width result in reaching the asymptotic ultimate regime  \cite{zhu18}, i.e.\ the $\Nuw\propto \Ta^{1/2}$ scaling is attained. 

\begin{figure}
\centering
\includegraphics[width=1\textwidth]{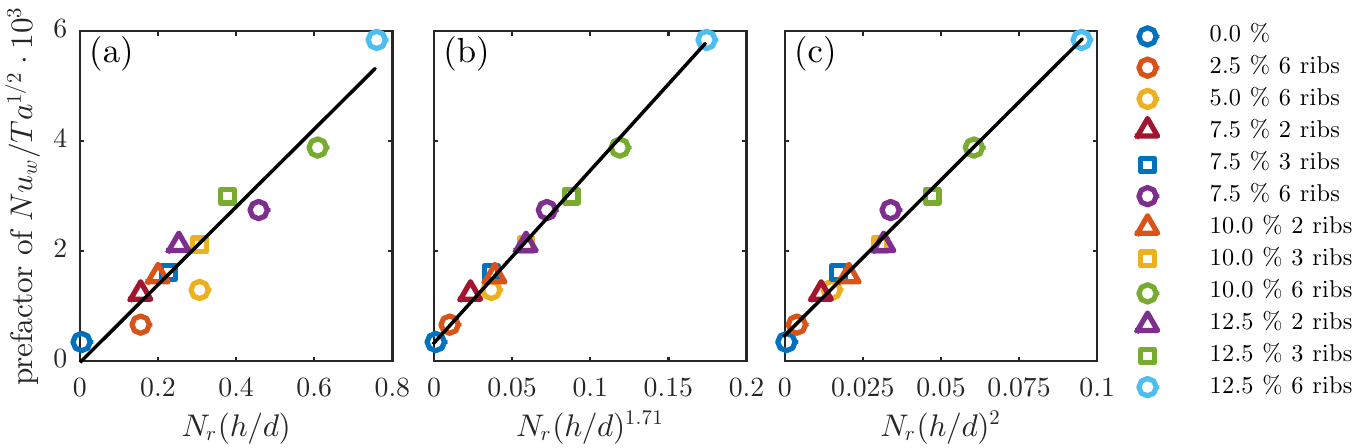}
\caption{Prefactor of the $\Nuw \propto \Ta^{1/2}$ scaling as a function of rib number $N_r$ and normalized rib height $h/d$, obtained from experiments. {\bf (a)} Results as a function of rib frontal area, which equals the $N_r(h/d)$. In {\bf (b)}, we show the best fit, showing that the prefactor scales with  $N_r  (h/d)^{1.71}$. In {\bf (c)}, we show the prefactor as a function of $N_r (h/d)^2$. The goodness of the fits is calculated here with the $R^2$ value. For figures (a), (b), and (c), they are $R^2_{b=1} = 0.9317 $, $R^2_{b=1.71} = 0.9953$, and $R^2_{b=2} = 0.9901 $, respectively.}
\label{Chap_Four_fig:fig3}
\end{figure}

To understand how the torque scales with rib height, we extract the mean prefactor of the experimental results  shown in figure \ref{Chap_Four_fig:fig2} over the measured Taylor number range, as shown in figure \ref{Chap_Four_fig:fig3}. We here plot experimental results for various roughness heights $h$ and rib number $N_r$. Intuitively, one could think that the prefactor scales with the total frontal area of the ribs, i.e.\ with $S=N_r h L $. When considering e.g. the drag equation $F_D = \frac12 \rho u_\infty^2 C_D S$, this argument indeed could be correct, as long as $C_D$ remains constant.
Although figure \ref{Chap_Four_fig:fig3}a indeed shows some correlation, the quality of the fit, which is shown as solid black line, can be improved. 
The fit quality can be increased by  assuming a scaling of type $N_r (h/d)^b$, in which $b$ is a fitting parameter. It is found that $b=1.71$ collapses our data in the best way. The $R^2$ values are given in de caption of figure \ref{Chap_Four_fig:fig3}.

\begin{figure}
\centering
\includegraphics[width=0.7 \textwidth]{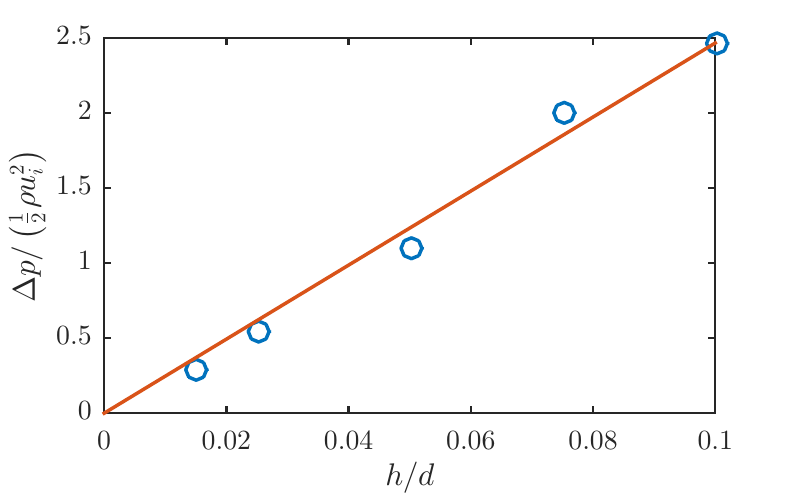}
\caption{Dimensionless pressure difference $\Delta p/\frac12 \rho u_i^2$ between upstream and downstream sides of the ribs as a function of rib height $h$ for $\Ta = 1 \times10^9$ and $a=0$, obtained from DNS. We observe a linear dependence between pressure difference and rib height. }
\label{Chap_Four_fig:fig4}
\end{figure}

\begin{figure}
\centering
\includegraphics[width=1 \textwidth]{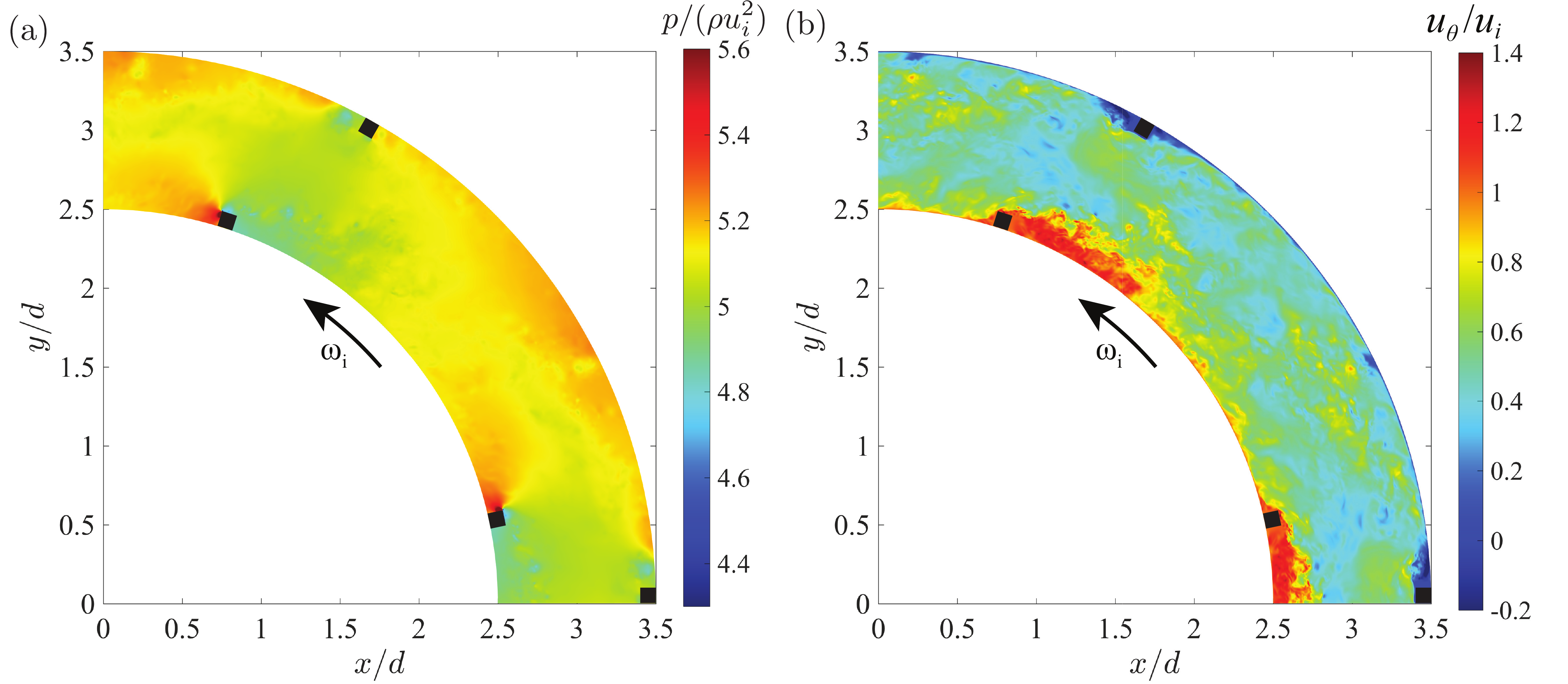}
\caption{{\bf (a)} Dimensionless pressure field, and {\bf (b)} dimensionless azimuthal velocity field obtained with DNS. The rib height is $0.1d$, and the Taylor number is $1\times 10^9$ and $a=0$. We here show one instantaneous field, taken at mid-height of the numerical domain. The inner cylinder rotates counter-clockwise as indicated, the outer cylinder is stationary. The positions of the ribs are indicated by the black squares.}
\label{Chap_Four_fig:fig5}
\end{figure}

As shown in ref.\ \cite{zhu18}, in the asymptotic ultimate regime, the pressure force results in the dominant injection of momentum, rather than the skin friction.  The torque $\tau_p$ exerted by the fluid on the inner cylinder through pressure forces on the ribs is given as:
\begin{equation}
\tau_p = r_i S \Delta p,
\end{equation}
in which  $S = h N_r L$ is the total frontal area of all the ribs, and $\Delta p$ is a mean pressure difference between the upstream and downstream side of the rib. An instantaneous pressure field is shown in figure \ref{Chap_Four_fig:fig5}a. Obviously, a local region of high pressure is found at the upstream side of the rib, and a local region of low pressure is present at the downstream side. In figure \ref{Chap_Four_fig:fig4}, we show the pressure difference $\Delta p$ as a function of rib height $h$. Indeed, the pressure difference is related to the rib height, and, is surprisingly well represented by the linear relation $\Delta p \propto h$.  With this knowledge we can now better understand the height dependence of the prefactor of the $\Nuw/\Ta^{1/2}$ relation: the pressure forces scale with the product frontal area $ N_r h L$, and pressure difference $\Delta p$, which is also proportional to the rib height $h$, leading to the prefactor to scale with $N_r (h/d)^2$, as long as the skin friction is negligible compared to the pressure forces, and in the case that ribs are unaffected by the neighbouring ribs. This result is close to what we found in our experiments, where we observe a slightly smaller scaling, namely the  aforementioned $N_r (h/d)^{1.71}$ scaling, presumably because the ribs cannot be regarded as isolated.
In addition, the analysis presented here is obviously limited to cases in which the pressure drag is dominant. Therefore, it does not fully cover cases with too sparse or dense rib densities, as well as too small rib heights \cite{zhu18}.

\section{Local results}
\label{sec:local}

\begin{figure}
\centering
\includegraphics[width=0.6\textwidth]{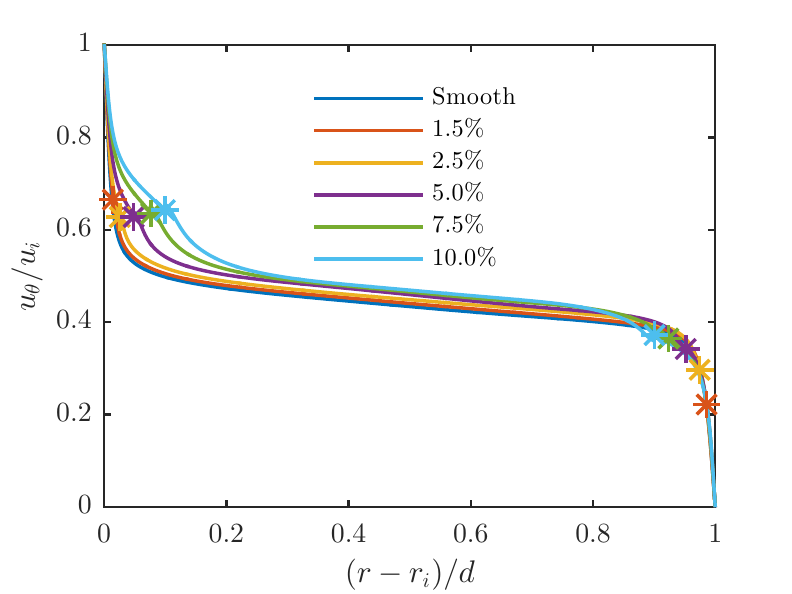}
\caption{Azimuthal velocity profiles for various roughness heights, non-dimensionalized by the inner cylinder velocity and the gap width at $\Ta = 1\times 10^9$, obtained from DNS. Here, six ribs are attached to both cylinders. The outer cylinder is kept stationary. The stars indicate the extent of the ribs.}
\label{Chap_Four_fig:fig6}
\end{figure}

\begin{figure}
\centering
\includegraphics[width=0.8\textwidth]{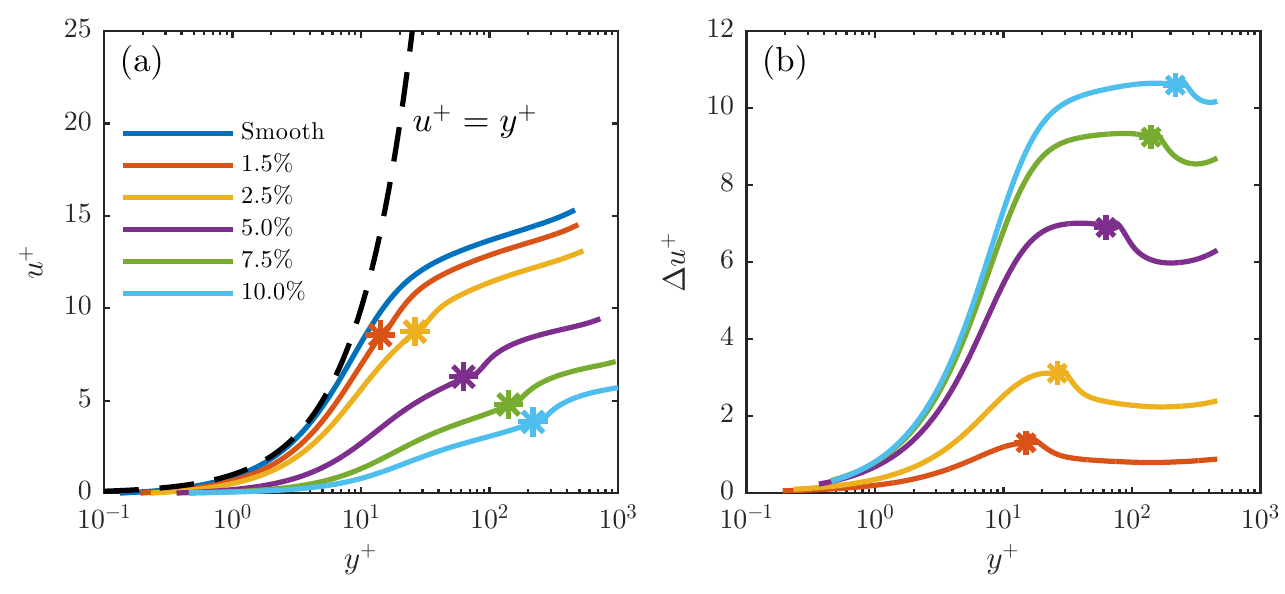}
\caption{{\bf (a)} Velocity profiles for various roughness heights non-dimensionalized by the friction velocity $u_{\tau}$ and wall distance to the inner cylinder $\delta_{\nu}$ at $\Ta = 1\times 10^9$, obtained from DNS. Here, six ribs are attached to both cylinders. The outer cylinder is kept stationary. The stars indicate the extent of the ribs. In {\bf (b)}, we show the difference from the smooth case, $\Delta u^+ = u^+_{smooth} - u^+$.}
\label{Chap_Four_fig:fig7}
\end{figure}

\begin{figure}
\centering
\includegraphics[width=0.8\textwidth]{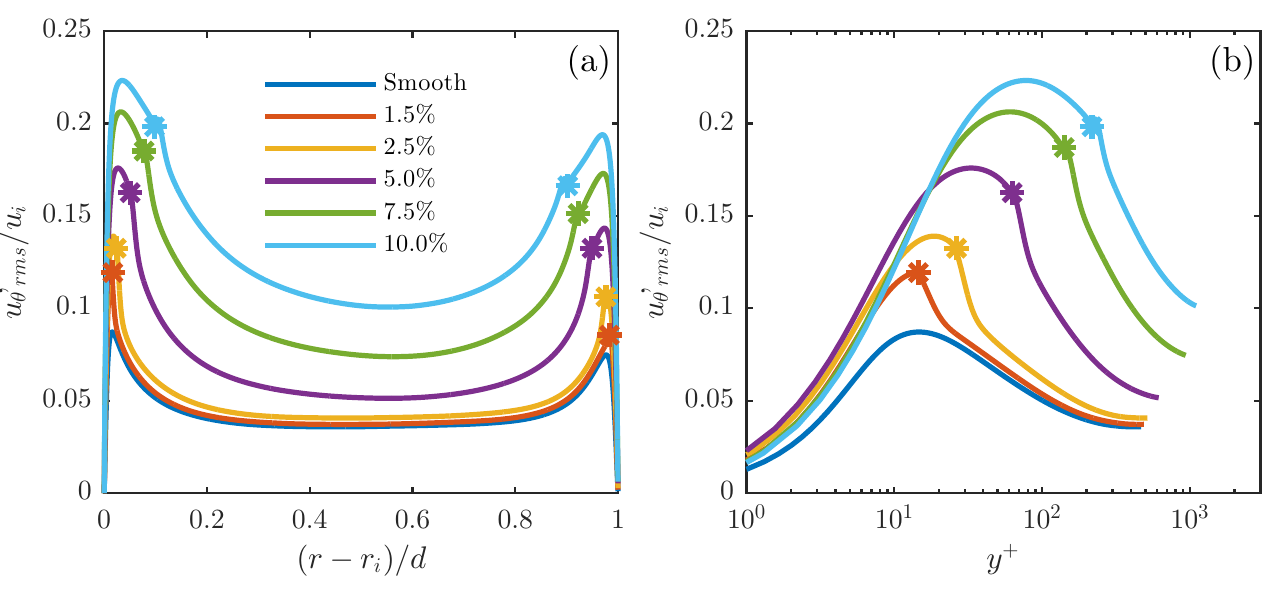}
\caption{RMS of the velocity fluctuations of the azimuthal velocity at $\Ta = 1\times 10^9$, obtained from DNS. Here, six ribs are attached to both cylinders. The outer cylinder is kept stationary. The stars indicate the extent of the ribs. In {\bf (a)}, the full fluctuation profiles are given, whereas in {\bf (b)} the fluctuations are shown on a semi-log scale as function of the wall distance to the inner cylinder.}
\label{Chap_Four_fig:fig8}
\end{figure}

We now show the azimuthal velocity profiles in figure \ref{Chap_Four_fig:fig6}, extracted from the DNS simulations which were shown in figure \ref{Chap_Four_fig:fig2}. 
Although the mean azimuthal velocity in the bulk remains largely unaffected by the roughness, that is not the case in the boundary layers (BLs), as one could expect for rough walls. Here we see that the BLs become thicker, and consequently that the wall-normal velocity gradient is less steep for all roughness cases. The difference in $u_{\theta}$ in the BLs for all roughness heights is difficult to observe in figure \ref{Chap_Four_fig:fig6}, and becomes clearer in figure \ref{Chap_Four_fig:fig7}.
In this figure, we show the azimuthal velocity profiles in the form $u^+ = (u_i -u_{\theta})/u_{\tau}$ as  function of the wall distance $y^+=y/\delta_{\nu}$ from the inner cylinder, in which the viscous lengthscale is calculated as $\delta_{\nu}= \nu / {u_{\tau}}$. We see, as is known for roughness, that $u^{+}$ decreases with roughness \cite{fla14}. One could think that rib roughness has a significantly different influence on the flow than e.g.\ sand grain roughness, as ribs act in a more local and isolated way than the uniform sand grain roughness. This is however not reflected in the data, neither in the global torque results, nor by these local velocity results, as  the results shown here are similar to a wide range of other types of roughness in other flow setups, see e.g.\ refs.\ \cite{jim04,Schultz2007,cha15,Thakkar2018}. Therefore our data suggests that the overall trends are independent to the type of roughness applied. 

We clearly see in both figures \ref{Chap_Four_fig:fig6} and \ref{Chap_Four_fig:fig7} that the shear at the cylinders becomes increasingly smaller with increasing roughness. As the skin friction is directly related to the shear at the wall, we see that this contribution to the torque gets smaller, wheareas the total torque is increased tremendously. 
This finding again confirms that skin friction is not the dominant way of injecting energy to the system, and again highlights the role of the pressure drag.
Velocity fluctuations (fig.\ \ref{Chap_Four_fig:fig8}) show a clear and large peak close to both cylinders, with velocities in the bulk being of around 5\% of the mean azimuthal velocity. As the rib size increases, the position of the aforementioned peaks is further away from the cylinder wall, the peak value however being at smaller $y^+$ values than the roughness height $h^+$, indicated in the figure by the asterisk symbols. This indicates that the equivalent sand roughness height $k_s$  of the roughness heights is significantly smaller than the rib height itself, as $k_s$ is expected to be located closer to the wall than the peak of the fluctuations.
\section{Optimal transport}
\label{sec:CR}

\begin{figure}
\centering
\includegraphics[width=0.8\textwidth]{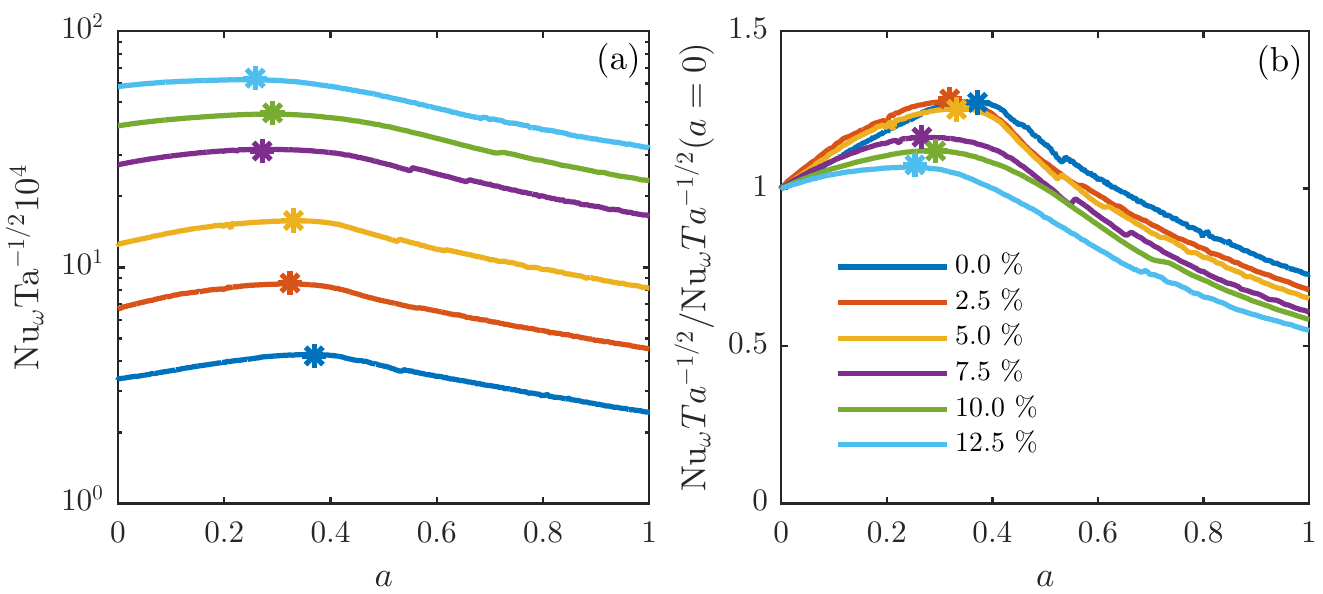}
\caption{{\bf (a)} The angular momentum transport $\Nuw$ as a function of rotation ratio $a$ and various rib heights (see legend) from the experiments. We fixed the number of ribs to 6. We multiply $\Nuw$ with $\Ta^{-0.5}$ to minimize temperature fluctuations in the experiments, which are reflected in the viscosity dependence of the Taylor number.  {\bf(b)} $\Nuw(a)$  normalized with its value for $a=0$, to highlight the differences between the shape of the curves. We indicated the maxima of each curve using an asterisk symbol.}
\label{Chap_Four_fig:optim_transp}
\end{figure}

So far we focussed on cases with stationary outer cylinder. In this section, we explore the behaviour of wall roughness in the counter-rotating regime.
For smooth walls, it is known that outer cylinder rotation has a significant influence on the momentum transport between both cylinders \cite{gil11,bra13,ost14pd}. Counter-rotating cylinders stimulate the existence of so-called ``turbulent Taylor vortices'', which enhance the momentum transport \cite{ost14pd,mar14}. At a rotation ratio of $a_{opt}=0.36$, the momentum transport reaches a maximum for the currently used radius ratio for the smooth wall case. In both extrema of $a= \pm \infty$, the flow (in the absence of end plates) is laminar, and thus $\Nuw=1$ \cite{pao11,gil12}.  Here we investigate what  the influence of ribs is on the behaviour in the counter-rotating regime. To study this, we fixed the Taylor number to $Ta = 3.8\times 10^{11}$, and  increased the rotation ratio $a$ from $a=0$ to $a=1$ in a quasi-stationary way.
As seen in figure \ref{Chap_Four_fig:optim_transp}, there is  a momentum  transport enhancement in the counter-rotating regime for all cases. When normalizing all curves by their $\Nuw(a=0)$ value, their shapes become very similar, i.e.\ apart from a prefactor, the behaviour is comparable. As in the smooth-wall case this `optimal transport peak' is related to the Taylor rolls, these results suggest that also in the presence of roughness Taylor rolls exist. As was shown in ref.\ \cite{zhu17}, ribs effectively shed off turbulent plumes, which  feed and drive the Taylor rolls. That even for extremely large roughness heights, here up to 12.5\% of the gap width, these rolls exist is surprising. The peak value of $\Nuw(a) / \Nuw(a=0)$ however does decrease for increasing rib height, indicating that the relative strength of the Taylor vortices decreases. 

\section{Conclusions and Outlook}
\label{sec:conc}
To conclude, building further upon our recently published work \cite{zhu18}, by providing further flow details on the local flow organization, the dependence on rib height and the behaviour in the regime of counter-rotating cylinders is illuminated.
We found that the momentum transport is largely enhanced with increasing rib height, caused by increasing pressure forces acting on the ribs. A scaling argument is found which predicts a scaling linear with rib number $N_r$ and squared with rib height $h$, i.e.\ $N_r (/d)h^2$. Experimentally, we find that to collapse the data the second  scaling exponent must be slightly smaller for the investigated $N_r$ and $h$, i.e.\ $N_r (h/d)^{1.71}$, presumably because the ribs cannot be regarded as isolated.
Velocity profiles and the near-wall velocities in wall units show that the velocity gradient close to the wall decreases with increasing roughness, similarly to what has been observed in other flow systems using other roughness types.

In the counter-rotating regime, the momentum transport depends on the rotation ratio similarly as in the smooth-wall case. Therefore, we hypothesize that in spite of the roughness, Taylor rolls still exist, and that their momentum-transporting role remains unaffected by it.  However, the ribs might decrease the effective radius, thus increasing the apparent aspect ratio and might thus allow for a larger number of rolls. Furthermore, the characteristics of the rolls can be affected by the significantly increased mixing. 
In addition, efforts are currently ongoing towards more realistic types of roughness, i.e.\ `sand-grain roughness', which are often encountered in turbulent flows encountered in engineering applications and in nature.

\begin{acknowledgments}
We would like to thank Gert-Wim Bruggert and Dennis van Gils for their  technical support over the years. We thank Pim Bullee, Rodrigo Ezeta and Pieter Berghout for various stimulating discussions. This work was financially supported by NWO-TTW, NWO-I, MCEC and a VIDI grant (No.\ 13477), which are all financed by the Netherlands Organisation for Scientific Research NWO, and the Natural Science Foundation of China under grant no.\ 11672156. The simulations were carried out on the national e-infrastructure of SURFsara, a subsidiary of SURF cooperation. We acknowledge PRACE for awarding us access to Marconi based in Italy at CINECA (No. 2016143351). We acknowledge that the results of this research have been achieved using the DECI resource Archer based in the United Kingdom at Edinburgh with support from PRACE under project 13DECI0246.

\end{acknowledgments}

\end{document}